# Hurricane Evacuation Analysis with Large-scale Mobile Device Location Data during Hurricane Ian

Luyu Liu [1, 2, *], Xiaojian Zhang [2], Shangkun Jiang [2], Xilei Zhao [2]

[1] Department of Geosciences, Auburn University, Alabama, USA

[2] Department of Civil and Coastal Engineering, University of Florida, Florida, USA

Hurricane Ian is the deadliest and costliest hurricane in Florida's history, with 2.5 million people ordered to evacuate. As we witness increasingly severe hurricanes in the context of climate change, mobile device location data offers an unprecedented opportunity to study hurricane evacuation behaviors. With a terabyte-level GPS dataset, we introduce a holistic hurricane evacuation behavior algorithm with a case study of Ian: we infer evacuees' departure time and categorize them into different behavioral groups, including self, voluntary, mandatory, shadow, and in-zone evacuees. Results show the landfall area (Fort Myers, Lee County) had lower out-of-zone but higher overall evacuation rate, while the predicted landfall area (Tampa, Hillsborough County) had the opposite, suggesting the effects of delayed evacuation order. Out-of-zone evacuation rates would increase from shore to inland. Spatiotemporal analysis identified three evacuation waves: during formation, before landfall, and after landfall. These insights are valuable for enhancing future disaster planning and management.

* Corresponding author: Luyu Liu, liuluyu0378@gmail.com, 0000-0002-6684-5570

**Highlights:**

- Terabyte-level GPS data reveals evacuation behaviors during Hurricane Ian in 2022.
- The landfall area, with late orders, had lower out-of-zone but higher overall evacuation rates.
- The predicted landfall area had higher full compliance rates, linked to early orders.
- Out-of-zone evacuation rates were higher for inland areas compared to coastal regions.
- Three main evacuation waves happened during formation, before landfall, and after landfall.

## 1. Introduction

The growing frequency, intensity, and uncertainties of disasters, especially hurricanes, have underscored the critical need to measure travel behaviors during evacuations. The escalating impact of climate change has also accentuated the urgency of this need, which makes large-scale evacuation events the new norm in the coastal areas. As one of the most vulnerable areas to climate change and extreme weather disasters in the United States, the State of Florida suffered disproportionately from the impacts of hurricanes. Florida's unique geography and demography also elevate hurricanes and evacuations to pivotal social equity issues, as vulnerable and marginalized communities often bear a disproportionate burden during such events. All these factors underscore the pressing need for understanding evacuation behavior.

Hurricane Ian in September 2022 was the costliest and the deadliest hurricane in Florida's history and the third-costliest weather disaster on record worldwide after Hurricane Katrina and Harvey (NOAA National Centers for Environmental Information, 2024). This deadly and extremely destructive Category 5 hurricane has caused more than 150 deaths in the US and $113 billion in damage (Bucci et al., 2023; NOAA National Centers for Environmental Information, 2024). Ahead of the landfall, Florida officials ordered the evacuation of 2.5 million people in 13 counties (Nyce, 2022). However, despite multiple efforts to issue evacuation orders at the federal, state, and county levels before landfall, numerous areas experienced notable delays in information release (Fleischer, 2022; Olivo et al., 2022). The prime example is Lee County, which accounts for almost half of the deaths, was reported to only issue evacuation orders not quite 36 hours before Ian made landfall (Fleischer, 2022). This raises numerous concerns towards the untimely responses of the county authorities, while many claimed that the latency in information release could contribute to the high number of fatalities (Olivo et al., 2022).

Despite the urgent need and demand to understand the behaviors of evacuees and their connections to different demographic factors, the enormity of the evacuation makes it difficult to precisely assess the detailed spatial and temporal patterns of evacuation behaviors. Traditional data collection methods, such as surveys and direct observations, are often not scalable nor timely enough to capture the complex dynamics of mass evacuations. Large-scale mobile device location data, a type of secondary data generated from smart devices with high-resolution spatiotemporal information of users' behaviors (Zhao et al., 2022), offer a new opportunity to study evacuation behavior in unprecedented detail. With tens of billions of location and time signals in this terabyte-level dataset, we can investigate the details of the evacuation behaviors in higher resolution and larger extent in both spatial and temporal dimension (Yabe et al., 2016; Yabe and Ukkusuri, 2020).

However, there are multiple major gaps existing in the application of GPS data in the hurricane evacuation studies. First, despite rich information contained in the GPS data, very few studies investigated the intricacies of evacuees' behaviors, such as evacuation timing and behaviors after leaving the home. Most prior GPS-based studies did not further classified the evacuees into different behavioral groups (Cheng, 2021; Tao, 2021; Washington et al., 2024), which is extremely crucial for the understanding of evacuees' travel behavior and decision making and future disaster planning (Wu et al., 2022; Zhao et al., 2022). Second, the application of large-scale disaggregated mobile device GPS is still lacking. Due to its nature of high heterogeneity and variety, it is still challenging to digest and translate the data into practical management and planning insights. Finally, as the one of the most destructive, deadliest, and costliest hurricanes in the US history, there are very few studies that focus on Hurricane Ian and investigate the evacuation behaviors during the event. Therefore, we have three research objectives in this paper as shown in Figure 1:

- **Data**: Create a robust data pipeline to digest and integrate terabyte-level mobile device location data in the context of evacuation behavioral studies.
- **Algorithm**: Develop a hurricane evacuation behavior algorithm to offer high-fidelity, high-resolution insights of evacuation behaviors.
- **Insight**: Measure the spatiotemporal patterns of evacuation during the Hurricane Ian, including the evacuation rate, different evacuation groups, and evacuation timing.

To meet these three goals, we develop a new hurricane evacuation behavior inference algorithm in the context of the Hurricane Ian, which categorizes the users into different evacuation types and infers their departure time. We also use a *proxy-home-location inference algorithm* (Zhang et al., 2024; Zhao et al., 2022) to infer the home location and the origin of each evacuation trip and a *user activity inference algorithm* to infer the activities and stay points of each users.

The paper is organized as follows. We first introduce the background of evacuation behavior studies and the gaps in current research in section 2. We then introduce our datasets and three algorithms we used in the paper in section 3. We continue to show the representativeness of the dataset and the results of spatiotemporal analysis in section 4. We conclude the paper with discussions on the insights we gain in the paper and the potential implications for future disaster research and strategic planning in section 5.

## 2. Background

The study of evacuation behavior involves various interdisciplinary methodologies spanning psychology, sociology, urban planning, and emergency management. In this section, we review the literature utilizing the mobile device location data to study disaster evacuation.

## 2.1. Evacuation Behavior Modeling with GPS Data

Various passively collected or secondary GPS datasets have been used to infer the evacuation behaviors in the context of disaster evacuation. These include mobile device location data (Washington et al., 2024), connected vehicle data (Ahmad et al., 2024), and social media analytics (Martín et al., 2020). A common trait of these methods is that the data are passively collected and not originally for the specific purposes of hurricane evacuation studies. This means that these data are usually timelier and more intensive in size but with less contextual information, such as trip purposes. Therefore, the modeling of human mobility and travel behavior is one of the most important scopes for disaster evacuation studies using GPS data.

Secondary GPS data are also widely employed to study emergency evacuations in the contexts of a variety of disasters, including hurricanes (Washington et al., 2024; Yabe and Ukkusuri, 2020), earthquakes (Horanont et al., 2013; Yabe et al., 2019), nuclear disaster (Hayano and Adachi, 2013), and wildfire (Cardil et al., 2019; Cova et al., 2024; Wang et al., 2017; Wu et al., 2022; Zhao et al., 2022). For instance, Hayano & Adachi (2013) employed GPS data to track the movement of people during the Fukushima Nuclear Power Plant Accident. Yabe et al. (2019) analyzed mobile phone location data to examine evacuation behaviors following earthquakes. In the context of wildfires, Wu et al. (2022) explored the wildfire evacuation decision molding with large-scale mobile device GPS data during the 2019 Kincade fire, and Cova et al. (2024) studied the wildfire evacuee trips' destination with the GPS data. Furthermore, Horanont et al. (2013) and Yabe et al. (2016) investigated how GPS data can be leveraged to analyze evacuation behavior in real time, which can provide decision-makers with crucial insights to optimize their emergency response efforts.

## 2.2. Hurricane Evacuation Inference Algorithm

Among the numerous data and case studies, a prime example is to infer the hurricane evacuation behaviors using GPS-based mobility tracking. This data allow researchers to create complex scenarios and infer individuals' behaviors under different conditions with large-scale empirical data in the context of population evacuation modeling, long-term recovery analysis, and inference of the damages to the built environment (Washington et al., 2024; Yabe et al., 2022). For GPS-based evacuation studies, there are two prime research questions: the inference of home locations and the determination of evacuation behaviors.

**Home Inference.** Most prior studies focus on the residents of the affected area and infer their evacuation by studying their mobility patterns outside their homes (Washington et al., 2024). Therefore, home detection algorithms (HDA) were widely used in evacuation studies; Verma et al. (2024) categorized HDAs into supervised, i.e., algorithms with validation of real home addresses, and unsupervised methods, i.e., algorithms that adopt assumptions to infer the home locations. Researchers have heavily replied on unsupervised HDAs due to the lack of high-fidelity home location information and privacy concerns (Verma et al., 2024). The most used assumptions are 1) places with more observed activities are more likely to be their home, and 2) users are more likely to stay home during nighttime and off-work time. The paper also compared the five existing HDA methods and found grid frequency method, which we use in this paper (discussed later), is among the methods with best performance.

**Evacuation Behaviors.** To determine whether a user if an evacuee during a hurricane event, the spatiotemporal constraint of the users activities is the most important factor, i.e., the time the user stays away from home and place staying (Washington et al., 2024). Washington et al. (2024) reviewed the different values of distance and time parameters utilized in prior studies. The time threshold to determine the evacuation behavior usually ranges from 1 days (Long et al., 2020; Younes et al., 2021) to 3 days (Cheng, 2021), with longer time threshold having the risk of underestimating shorter duration evacuation (Lindell et al., 2011; Tao, 2021). Another major parameter is the threshold of home departure activity, which can range from 50 meters (Deng et al., 2021) to 1600 meters (Darzi et al., 2021); this will also significantly impact the outcome of the inference process. Based on these prior studies, we choose the values for these two parameters in our evacuation inference algorithm (more discussed below). Beyond the simplistic spatiotemporal constraints, some studies also applied more refined rule-based assumptions, especially the introduction of nighttime stay strategy, i.e., how long a user would stay outside the home during nights (Yabe et al., 2019). Similar to the two assumptions adopted by the home detection algorithms, this rule helps to improve the performance of the algorithm and the details of the results.

2.3. Research gaps

However, there are still multiple research gaps in this domain. First, prior studies do not address the fine-grained behavior throughout evacuation trips in the context of hurricanes, especially after the users left their homes. This is especially important considering the longer duration of the disaster. With more abundant information in the large-scale GPS data, we can create a more detailed and accurate representation of evacuees' behaviors, leading to greater accuracy and authenticity. Prior studies also did not further classify the affected population into multiple behavioral groups based on their departure time, origin, and destination. These findings would greatly help scientific and planning communities and authorities to better understand the evacuation patterns and improve decision-making processes and planning strategies (Wu et al., 2022; Zhao et al., 2022).

Empirically, despite multiple studies on prior hurricanes with disaggregated GPS data (Cheng, 2021; Tao, 2021; Washington et al., 2024), there are very few studies that focus on Hurricane Ian. To the best of our knowledge, only three papers discuss the evacuation behavior during the Hurricane Ian. Li et al. (2024) used aggregated Streetlight OD matrix data to measure the mobility change in different counties during Ian. Ebrahim et al. (2024) used traffic data to calculate the evacuation road utilization during Ian. Alam et al. (2024) conducted a Pollfish survey (n=100) to study people's safety perceptions during Ian. This study will address these research gaps and provide comprehensive insights by exploring previously overlooked aspects, revealing nuanced patterns, and suggesting strategies to shape future research.

## 3. Methods

In this section, we introduce a systematic overview of the data, algorithms, and analyses conducted in this paper in the context of Hurricane Ian as shown in Figure 1. Each layer corresponds to a research objective in the introduction section from data to algorithm and then to insights, which we extensively discussed below in each section.

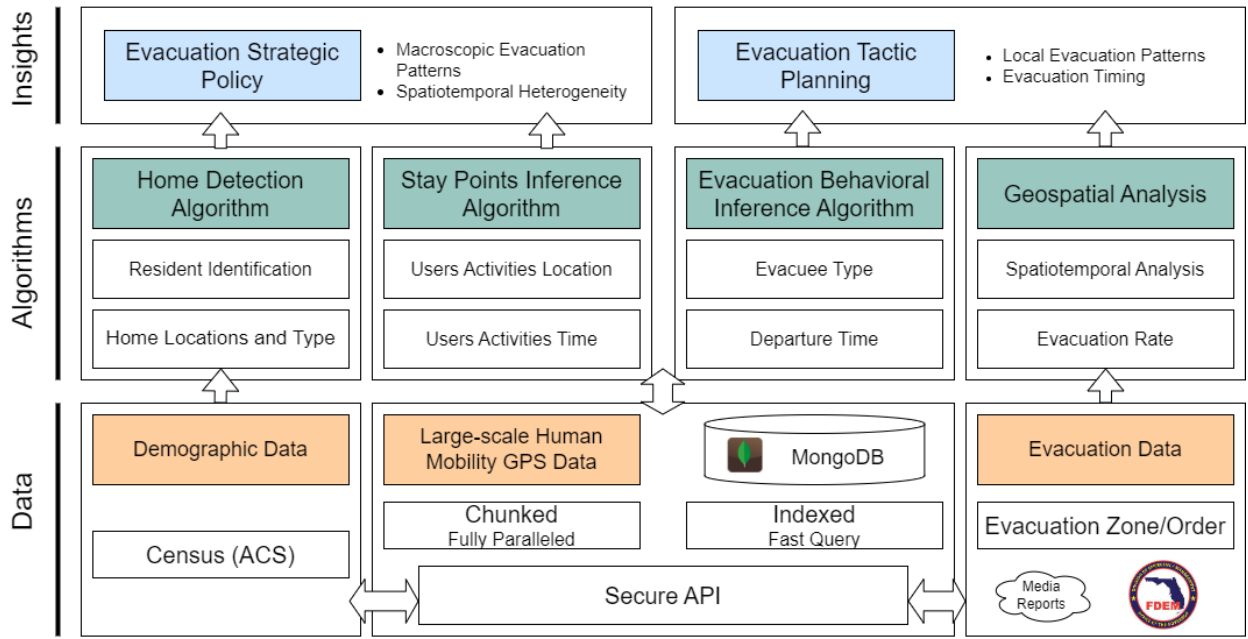

Figure 1: The structure of the evacuation behaviors modeling.

3.1. Data

As we show in the first layer of Figure 1, we used three major datasets: large-scale mobile device location data, evacuation order data, demographic data. The primary data source we use is the mobile device location data collected by Gravy Analytics, a location intelligence company focusing on human mobility data (Gravy Analytics, 2024; Zhao et al., 2022). This large-scale, high-resolution, and disaggregated GPS dataset contains over 150 million mobile devices in the United States, with 13 billion generated records and 2.6 terabyte in data size from Sep 1 to Oct 15, 2022 in Florida. The information provided in the GPS data includes location, timestamp, forensic flag (i.e., the GPS errors resulting from factors such as spoofed locations, cell tower and IP generated signals, abnormal signal density, and GPS signal floating).

It is a major challenge to filter and sort such a large dataset. We first insert the raw dataset into a MongoDB database, a nonrelational NoSQL database, for its flexibility to handle large-scale data, support for real-time analysis, and high-speed data ingestion (Banker et al., 2016). In Python and MongoDB environment, we employ data cleaning techniques on the raw dataset. We first merge and sort the data based on user ID and timestamp and create indexes for the two fields. To ensure the quality of the data and remove users with low frequent data, we filter out duplicate points and all users with less than 150 GPS points and only keep the medium-high to high accuracy points defined by Gravy, where the GPS errors are no more than 50 meters (Gravy Analytics, 2023). We finally project all the coordinate points to WGS 84 / UTM zone 17N projection system (EPSG:32617), whose unit is meters. All steps are fully paralleled with the multiprocessing package in Python environment to ensure performance.

We visualize the patterns of the sampling rate in Figure 5**Error! Reference source not found.**. The sampling rate is calculated as the ratio between the sum of GPS-inferred evacuees and non-evacuees and the total population from ACS 2018 – 2022 5-year estimates. The average sampling rate of users are 5.03% (standard deviation = 2.76%), with 250939 residents from GPS and 4990438 residents from ACS data. It shows that the regional differences between different areas are not very drastic; most areas with low sampling rate are located near the outskirts of the evacuation zone, which is the result of inconsistent polygon overlaps between census tracts and evacuation zone.

Our case study focuses on Hurricane Ian, the third costliest and second deadliest hurricanes in the United State history. We collect all the information about the evacuation orders, including releasing time, evacuation zones, and order changes over time as shown in Figure 2. We focus on the west coast of Florida, where Hurricane Ian primarily impacted during its landfall. Our study area encompasses all Florida counties that issued an evacuation order, either voluntary or mandatory. We collect and cross-validate the times and levels of evacuation orders from each county's official order statement or social media platforms and multiple media reports (McKernan, 2022; Victor and Chung, 2022). We also generate a 7.5 km buffer zone of the evacuation zones and infer evacuation shadow evacuation behaviors in those places.

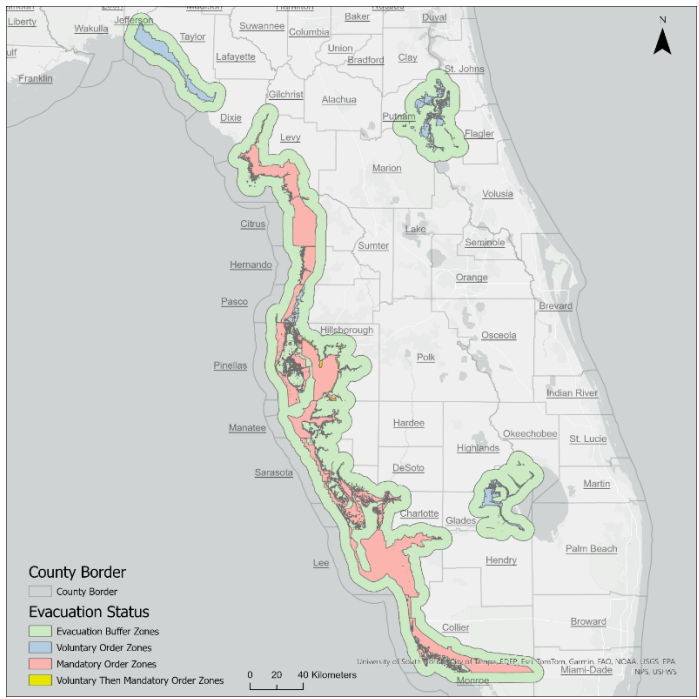

Figure 2: the map of case study site. Red color represents areas with mandatory order zone, blue color represents areas with voluntary order zone, green color represents the buffer zone, and yellow color represents voluntary order zone that issued a mandatory order later.

3.2. Home Detection Algorithm

We used the home detection algorithm introduced in Zhao et al. (2022), which is among the best home detection algorithms according to Verma et al. (2024). It is the first step to understand evacuation behaviors by determining the origin of the trips. In this study, we assume that all evacuees started their evacuation trips from home, which is a reasonable assumption widely adopted by prior studies (Zhao et al., 2022). Figure 3 shows an example to infer proxy-home location for each user in the GPS data. We first fit all the GPS points in a 20-meter-by-20-meter grid, which is a typical home lot size in Florida, and find the stay duration for each cell. Based on the duration of the points in each cell, we adopt a parallel approach to infer users' home from points generated during nights and during weekends, respectively. We first find the cell that the user stayed for the longest duration during each night, and we assign the most stayed cell as the proxy home location if the user spent at least 5 nights at the cell. We choose 5 nights as a threshold for residents because tourists stay in Florida for 4.3 days in average (Harrington et al., 2017). If no night GPS points were found, we will identify the cell that is most frequently stayed during weekend, and we assign the most stayed cell as home if users spent at least 6 hours during weekend at the cell. We choose 6 hours as the threshold because Americans spend at least six hours conducting activities at home per the American Time Use Survey 2022 (Bureau of Labor Statistics, 2023). After the inference, we further filter out those users who have less than 15 days with activities (each day must have at least 10 GPS points) during the whole lifespan. Note that the residents' home locations are inferred with data before hurricane Ian.

3.3. Stay Points Inference Algorithm

We use the stay points inference algorithm introduced by Zhang et al. (2024) to infer the activities of each user. An activity or a stay point is defined as a cluster of GPS points that are considered to happen during a single event finished by a user. The purpose is to contextualize the mobility data and reduce redundancy. For example, we would like to merge large amounts of GPS points clustered in the same place into a single activity at different times. Another example is to remove fly-over points generated on the highway that happened in a very short span, which cannot be regarded as an individual activity or stay point.

In practice, we first get each user's GPS points sorted by their timestamp. We add the first GPS point to a cluster and iterate each GPS point and check the eligibility of each point to be added to the cluster by applying the two spatiotemporal constraints. The GPS point will be added to the cluster if the GPS point is less than 100 meters from the center of the cluster (Chen et al., 2014). If further than 100 meters, then we check if the duration of the cluster is longer than 5 minutes (Wang et al., 2019; Zhang et al., 2023); if not, then disqualify the cluster as fly-over points. Each qualified activity contains the location, i.e., the centroid of the cluster, and the start, the duration, and the end of the event.

A major advantage of this algorithm is that it considers the duration of time, rather than count of GPS points, as the primary indicator of presence. As mobile device location data can have much redundancy and highly uneven temporal density for different users and time, this strategy significantly decreases the behavior bias of the results. However, due to the potential inherent sampling bias in the data, there can be still a long gap between GPS points; the activities that fall within the gap cannot be inferred by the algorithm, which were not recorded by the data. In that sense, we address the disproportionately higher density of points but cannot address the disproportionately lower density of points due to behavioral bias.

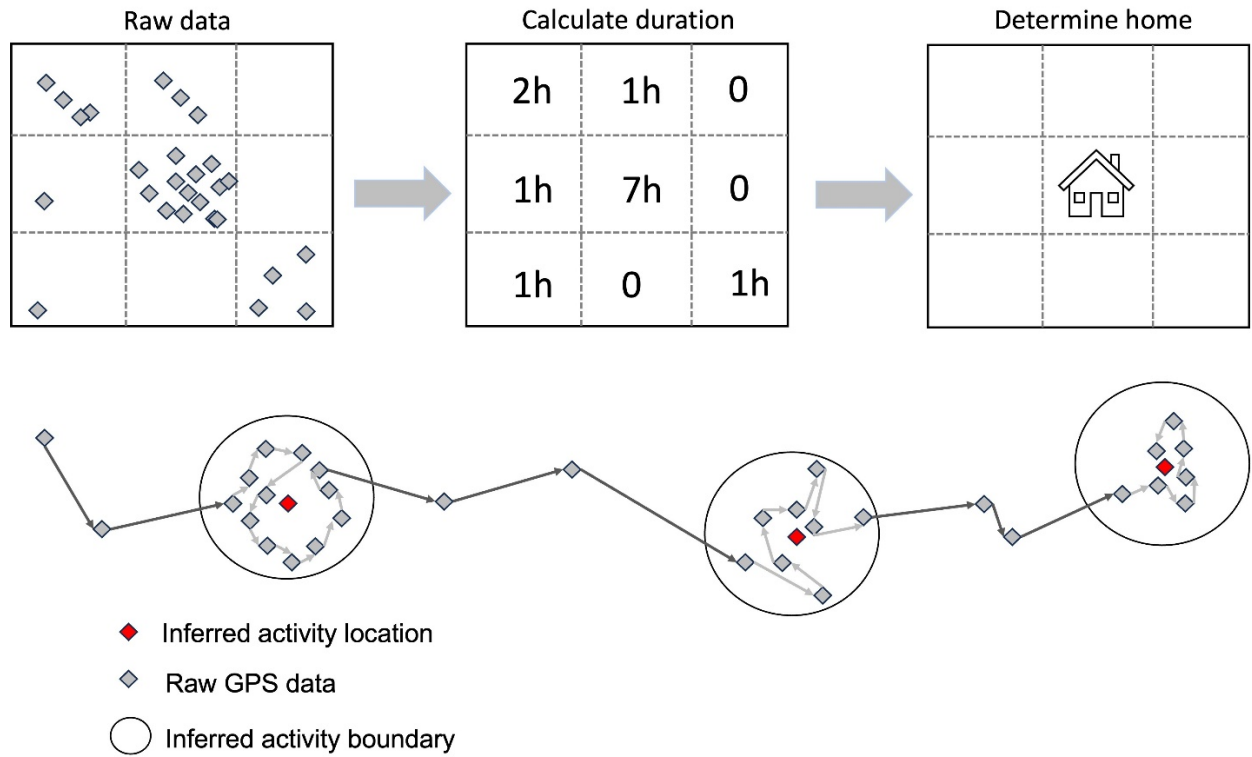

Figure 3: A simple demonstration of home detection algorithm (up) and stay point inference algorithm (bottom).

3.4. Evacuation Behavior Inference

Based on the users' home location and activity points, we introduce a new evacuation behavioral inference algorithm as shown in Figure 4. The purpose is to categorize users into different evacuation groups and then infer the evacuation time for evacuees. We have seven groups in this paper: *non-evacuees, shadow evacuees, self-evacuees, voluntary evacuees, mandatory evacuees, in-zone evacuees,* and *uncategorized personnels*.

- Non-evacuees, who did not evacuate. They stayed at home for most of the time during the hurricane (Wong et al., 2018; Younes et al., 2021; Zhao et al., 2022).
- Shadow evacuees, who evacuated after the issue of a nearby evacuation order but did not live inside the evacuation zones. This is a concept widely used in disaster response studies such as hurricane (Gladwin and Peacock, 2012), nuclear disaster (Zeigler et al., 1981), and wildfire (Zhao et al., 2022). The presence of shadow evacuees could have negative impacts on the evacuation of people ordered to evacuate, causing delays or traffic congestions along major evacuation routes (Tanim et al., 2022).
- Self-evacuees, who evacuated before any official orders (Cova et al., 2024; Zhao et al., 2022).
- Voluntary ordered evacuees, who evacuated after a voluntary order. This is similar to the evacuees under warning in prior wildfire evacuation studies (Cova et al., 2024; Zhao et al., 2022).
- Mandatory ordered evacuees, who evacuated after a mandatory order (Wong et al., 2018; Younes et al., 2021; Zhao et al., 2022).
- In-zone evacuees, who evacuated but still had activities inside the evacuation zones (voluntary and mandatory zone). Some individuals may comply partially with evacuation orders, moving to safer but closer locations within the evacuation zone instead of leaving the zone entirely. In other words, despite leaving their homes, these in-zone evacuees did not fully leave the evacuation zones. This can be due to multiple factors, such as resource access, risk perception, and logistical constraints (Wong et al., 2018; Younes et al., 2021). While out-of-zone evacuation behavior is extensively discussed, very few studies have studied the implication of partially compliant evacuees or in-zone evacuation behavior, which motivates us to identify patterns of partial compliance and explore the factors that influence individuals to remain within evacuation zones.

- Uncategorized persons, whose status cannot be inferred based on the information in the GPS data (Zhao et al., 2022).

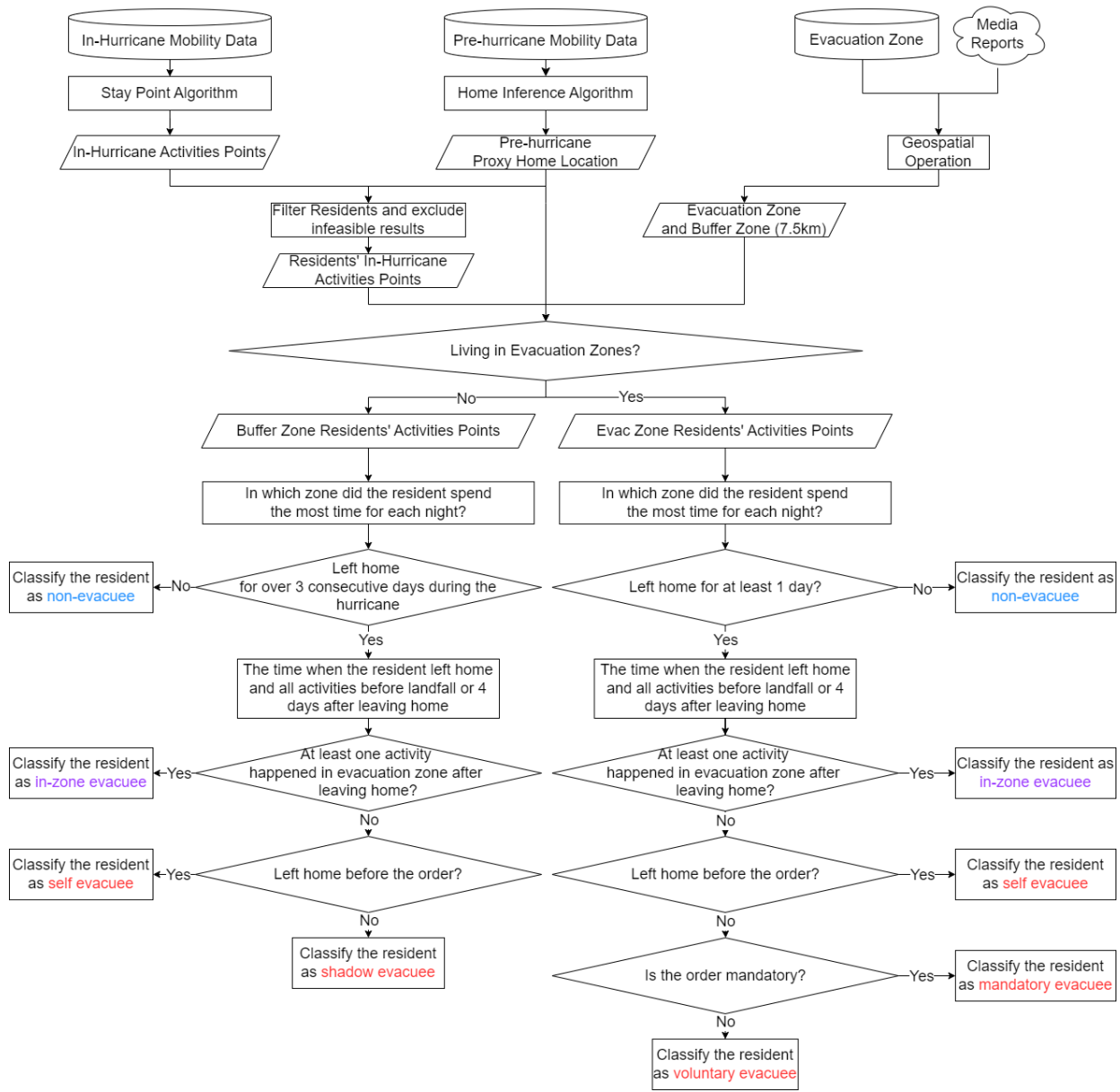

Figure 4: the evacuation behavior inference algorithm.

The time range of evacuation behaviors inference algorithm is from September 23rd to September 29th (Bucci et al., 2023). We retrieve all activities (i.e., the places that the users stayed for at least 5 minutes) during the time range. Based on the duration of each activity, we generate each user's mobility pattern in different zones. We find the time spent in home, buffer zones, voluntary zones, mandatory zones, and all other places; we then find the zone the user spent the most for each night from Sep 23rd to 29th.

The second step is data imputation. Not every day has activities. Therefore, if there is a day without any activities, which also means that we cannot find the place that the user stayed for the longest period during the day, we designate the most frequently visited zone on the nearest subsequent day. We tag the users as *uncategorized personnels* if days without data are longer than 3 days (50% of hurricane period). The third step is to find the departure time for evacuees. We assign the end time of the last activity that the user stayed at home as the user's departure time. If no time is found, which means the user did not stay at home at all, we assign departure time as the first activity's start time.

The final and most critical step is to categorize users into different groups. We adopted two different approaches to infer the evacuation behaviors in buffer zones and evacuation zones (including both voluntary and mandatory zones). For users living inside the evacuation order zones, we categorize the users living in evacuation zones based on two criteria: 1) if the user left home for at least 1 night between 23rd and 29th; 2) if there were no activities inside the evacuation zones until the landfall time (3:05 PM Eastern Time, September 28, 2022) or four days after the home departure time if evacuating after the landfall. If qualifying for both criteria, we further compare the departure time with the time of evacuation order. We compare the home departure time with the order time: if the departure time is earlier than the order time, then assign the user as a *self-evacuee*; if the departure time is later than or the same as the order time, then assign the user as an *ordered-evacuee* or *a voluntary evacuee*, depending on the type of the evacuation order. If the user only qualifies the first condition, which means the user left home but still have stay points inside the evacuation zones, we then categorize the user as *in-zone evacuee*.

Likewise, for users living in the buffer zone, we identify evacuees based on two criteria: 1) if the user left home for at least 3 nights and does not stay at home during all nights between 23$^{rd}$ and 29$^{th}$; 2) if there were no activities inside the evacuation zones until the landfall time or four days after home departure time if evacuating after the landfall. If qualifying for both criteria, we further compare the departure time with the time of evacuation order. We assign a buffer zone's order time based on its closest evacuation zone's issue time. We then compare the home departure time with the order time: if the departure time is earlier than the order time, then assign the user as a *self-evacuee*; if the departure time is later than or the same as the order time, then assign the user as a *shadow-evacuee*. Similarly, if the user only qualifies the first condition, which means the user left home but still have activities insides the evacuation zones, we then categorize the user as *in-zone evacuee*.

3.5. Analysis

We first visualize and analyze the representativeness of the large-scale GPS data, as prior studies have reported the potential mobility bias issues for GPS data generated from mobile devices (Harrison et al., 2020; Wang et al., 2019). We visualize the spatial distribution of the sampling rate in the study area, and we also provide a scatter point graph between the number of inferred residents and population from American Community Survey 5-year estimate 2018 – 2022.

Based on the algorithms introduced above, we further conduct several spatiotemporal analyses to investigate the different perspectives of the evacuation behaviors during the hurricane. We first categorize all users into the seven evacuation groups we discussed above and calculate the *evacuation rate,* i.e., ratio of shadow-evacuees, self-evacuees, voluntary evacuees, ordered evacuees, and in-zone evacuees out of all users except uncategorized persons. We also calculated the *out-of-zone evacuation rate*, i.e., the ratio of evacuees without in-zone evacuees, as a stricter measurement of evacuation level. Moreover, we conduct descriptive analyses of evacuation behaviors, including the overall, spatial, and temporal patterns of the evacuation rate. We also plot the evacuation response curves of all evacuees and different evacuee groups.

## 4. Results

Per the methods introduced above, we present our findings organized in three sections: 1) data representativeness; 2) spatial analyses; 3) temporal analysis.

### 4.1. Data Representativeness

Figure 5 maps the sampling rate in each census block group in the affected areas, i.e., the ratio of inferred residents from the home detection algorithm and population from ACS 2018 – 2022 5-year estimate data. The map uses a natural breaks (Jenks) classification. First, the map shows that the sampling rates are not drastically different among different areas other than very few outliners, such as the Everglades National Park and a few high-density neighborhoods in downtown Tampa. We also visualize a scatter point plot to show the relationship between inferred residents and population for all census block groups. The R is 0.69, which is close to the performance of census block group-level aggregated mobile device location datasets reported by prior studies (Jardel and Delamater, 2024; Z. Li et al., 2024).

The two graphs suggest a satisfactory level of representativeness across most of the study area, indicating that the mobile GPS data used captures a proportional snapshot of the broader population dynamics within these regions. This alignment supports the reliability of the findings derived from the GPS data and strengthens the confidence in using this approach for detailed evacuation studies.

Furthermore, despite showing a strong correlation, we do observe some outliners. In general, urban areas have lower sampling rate, and rural areas have higher sampling rate, which is consistent with prior conclusions (Z. Li et al., 2024). For some coastal neighborhoods, due to geofences guided by privacy or administrative purposes, these neighborhoods are not included in the GPS data. These places include the Sanibel Island and some portions of Pine Island in Lee County, which are among the most affected places during Ian. In the practical sense, the results drawn from this GPS dataset may underestimate or even miss the evacuation behaviors in those areas. We further discuss the implications of this issue in the conclusion section.

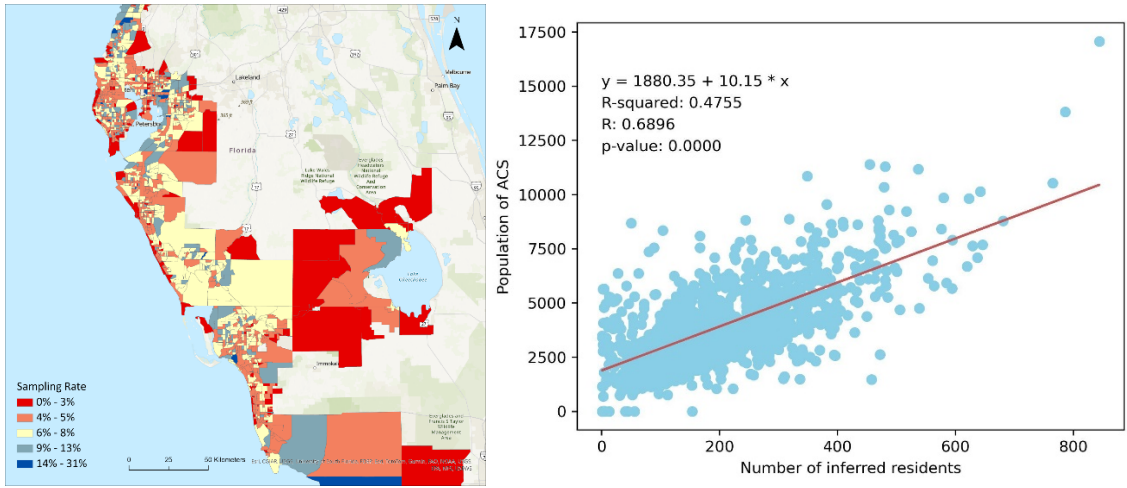

Figure 5: the map of sampling rate and the scatter point plot of population and inferred residents for each census block group in the affected areas.

4.2. Spatial Analyses

Table 1 shows the user counts of 6 user groups for Lee County (major cities: Cape Coral and Fort Myers), Hillsborough County (major cities: Tampa and Plant City), and all the affected counties in Florida. The table also shows the out-of-zone evacuation rate, which measures the number of residents who were fully in compliant with the evacuation orders and left the affected zones, and evacuation rate, which also counts the in-zone evacuees that did evacuated home but not fully follow the orders.

Table 1: Details of each evacuee group and the evacuation rates for all affected counties in Florida, Lee County (Fort Myers and Cape Coral), and Hillsborough County (Tampa).

| | Uncategorized | Non-evacuee | Shadow evacuee | Self-evacuee | Voluntary evacuee | Ordered evacuee | In-zone evacuee | Out-of-zone evacuation rate | Overall evacuation rate |
|---|---|---|---|---|---|---|---|---|---|

| | | | | | | | | |
|---|---|---|---|---|---|---|---|---|
| **Florida** | 43063 | 159335 | 16793 | 19782 | 245 | 5358 | 49426 | 14.3% | 31.2% |
| **Lee** | 6568 | 21254 | 810 | 2141 | 0 | 1058 | 12252 | 9.1% | 36.9% |
| **Hillsborough** | 11319 | 42214 | 7297 | 5277 | 245 | 1144 | 8185 | 18.4% | 29.3% |

First, the evacuation rate during the hurricane Ian is much lower than prior hurricanes and studies, which is consistent with the existing studies and media reports. Peace et al. (2022) reported that the authorities at Pinellas County expected 35,000 to evacuated to county-run shelter with fewer than 6,000 people showed up, and Hillsborough only had less than half of the expected number, which were much less than the numbers during the Hurricane Irma. As a reference, the evacuation rate reached 74% inside the evacuation zone in Jacksonville, Florida during Hurricane Matthew in 2016 (Ling et al., 2021), and the evacuation rate in the landfall area was reported to reached 90% during Hurricane Irma in 2017, primarily due to the effective door-to-door warnings by local police officers (Feng and Lin, 2021). For the case of Tampa, Tanim et al. (2023) reported an evacuation rate of 29.09 % in Hillsborough County during Hurricane Irma, and Martín et al. (2020) used Twitter and survey data and reported the evacuation rate as 28.4%. Considering that Tampa was predicted as the landfall site and given a mandatory evacuation order in those area, the evacuation rate during hurricane Ian is still low even when including in-zone evacuees.

Moreover, we also observe a highly heterogenous spatial pattern of evacuation rates in different counties. Ironically, as the landfall site with more than half of the casualties, Lee County has a very low out-of-zone evacuation rate compared with Hillsborough County and the global average, and the evacuation rate, including in-zone evacuees, is actually higher than Hillsborough County and the global average. Figure 6, moreover, visualizes the spatial pattern of the evacuation rates in each census block group. There are very few residents choosing to evacuate according to the evacuation order, and most people who left home did not fully evacuate out of the affected zone. This is consistent with the conclusion from prior studies that most surveyed evacuees from Lee County evacuated to the home of a friend or relative (Alam et al., 2024). This also exemplifies a trend of incompliance among the public when facing a major disaster in recent years (Connolly et al., 2020). Communication challenges and confusion about the zones may contribute to the low compliance rates. Furthermore, trust in the authorities and previous experiences with evacuations can also influence how people react to orders (Lazo et al., 2015).

In the context of Ian, these patterns can be explained by two factors: *prediction uncertainties* and *latency in evacuation orders*. Figure 6 visualizes two predictive trajectories provided by NOAA at 2am and 8am on Sept 27, less than 36 hours ahead of the landfall (3pm on Sept 28) (NOAA National Centers for Environmental Information, 2024). The projected landfall site was still in Tampa until 2pm Sept 27, which could encourage more people to evacuate and contribute to the high evacuation rate in Hillsborough County. Meanwhile, local authority in Hillsborough and nearby Pinellas County released a relatively early mandatory evacuation order on Sept 26, which gave the local residents more time to fully prepare and evacuate. This can be the reason why the ratio of out-of-zone evacuees is much higher than Lee and the global average.

On the other hand, authorities in Lee County did not release any evacuation orders until 7am Sept 27 (Fleischer, 2022; Lee County, 2022; Victor and Chung, 2022), citing the uncertainties in predictions (Olivo et al., 2022). As Connolly et al. (2020) pointed out, the indecisiveness and delay in releasing the evacuation orders can have a negative impact on people's willingness to comply. This decision also left residents who are willing to evacuate having less than 36 hours to prepare for the evacuation until the landfall. This is also consistent with the fact that there are higher numbers of in-zone evacuees possibly due to the lack of time (discussed below in Figure 8).

Shadow evacuation behavior is also highly heterogeneous across different areas. With more time and widespread evacuation order, the shadow evacuation behavior in Hillsborough County is more prevalent, while Lee County witnessed little to none shadow evacuation (also shown in Figure 7). Although shadow evacuation behaviors are not desirable in the context of mass evacuation planning and administration, this indeed reveals a lower willingness and motivation to evacuate in Lee County, possibly due to the lack of prompt evacuation order and prediction uncertainties.

Figure 6 also reveals an interesting phenomenon that the evacuation rate generally decreases from shore to inland, while out-of-zone evacuation rate increases from shore to inland. Prior studies has extensively discussed and confirmed that evacuation rate among coastal populations is always higher than inland populations (Mongold et al., 2021; Washington et al., 2024), which is consistent with the findings of our paper. However, if investigating the evacuation behaviors with more details and stricter criteria, this conclusion may not hold true for evacuees during Ian who fully complied with the evacuation orders. One of the major reasons is that coastal population are harder and have to travel longer distance to escape outside the designated zone, especially given the condition where there is little preparation time for longer-distance evacuation as we discussed earlier.

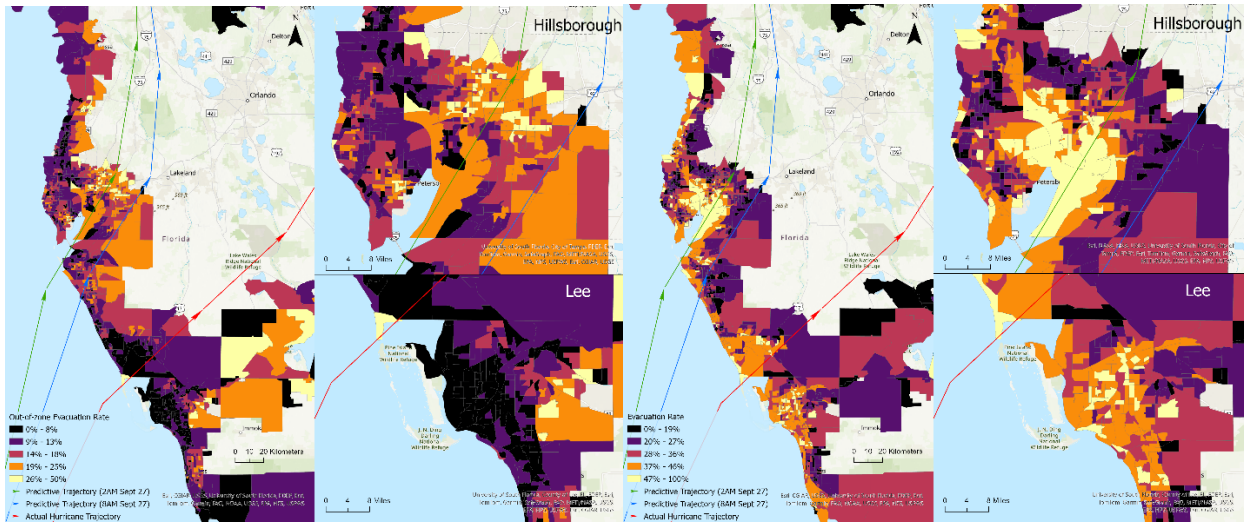

Figure 6: Out-of-zone evacuation rate and evacuation rate of each census block affected in the west coast of Florida, Hillsborough County (Tampa), and Lee County (Fort Myers and Cape Coral). Green and blue arrows represent the projected trajectory of hurricane Ian given by NOAA at 2am and 8am Sept 27, and the red arrows are the actual trajectory. All maps are generated with the Natural Breaks (Jenks) classification.

4.3. Temporal Analysis

Figure 7 visualizes the hourly pattern of evacuation behavior in all affected areas in Florida, Hillsborough County, and Lee County. All three graphs have three major high clusters of evacuation behaviors: first wave (12am 9/23 – 12am 9/24) – the early evacuees, second wave (12pm 9/26 – 12pm 9/27) – the evacuees near the landfall, and third wave (after 12pm 9/29) – the late evacuees.

The first cluster is from 9/23 – 9/24 when the hurricane was coming into form. This smaller cluster would be due to more sensitive and cautious evacuees who left their home early. The second cluster is near 9/27 just before the landfall, which can be accounted by the evacuees under the issued evacuation orders. After the second cluster, it is noteworthy that there is a valley in the curves from 12pm 9/27 and 12am 9/29, which falls into the hurricane landfall period (3pm 9/28) and can be because people would stay home and avoid the bad weather. Finally, soon after the hurricane, we observed a rapid increase in the number of evacuees for all groups. Note that the third wave evacuation is especially more significant in Lee County, which can be due to people escaping the flood. Per the tides and currents data (NOAA, 2024), the water level in Fort Myers (Lee County) reached the highest level of 2.5 meters at 5:30 pm 9/28, which indicates that the city is largely flooded. The water level decreased to 1.2 meters at 7am 9/29 and later decreased to 0.4 meters, the pre-hurricane level, at 12am 9/30.

Meanwhile, it is noteworthy that the three graphs share similar profiles but have different composition. The similarity between the two graphs suggests that despite differences in evacuation orders and spatial patterns, the evacuation process in different regions is largely synchronized. In that sense, the effect of evacuation orders is not strong enough to fundamentally alter the temporal patterns in different counties, since county authorities, who are responsible to release evacuation orders, are not the only information source for residents. However, we do also observe the significant differences in the composition of the evacuees. For example, Lee County has much lower level of evacuation before the landfall and evacuation order. Lee County's self-evacuees account for 4.8% of all inferred residents, while Hillsborough County's self-evacuees account for 7%, which indeed suggests the effect of evacuation order and trajectory prediction. In conclusion, our findings suggest that while the effects of evacuation orders are not substantial enough to make the temporal curve different in various counties, evacuation orders do impact the behavior of evacuees.

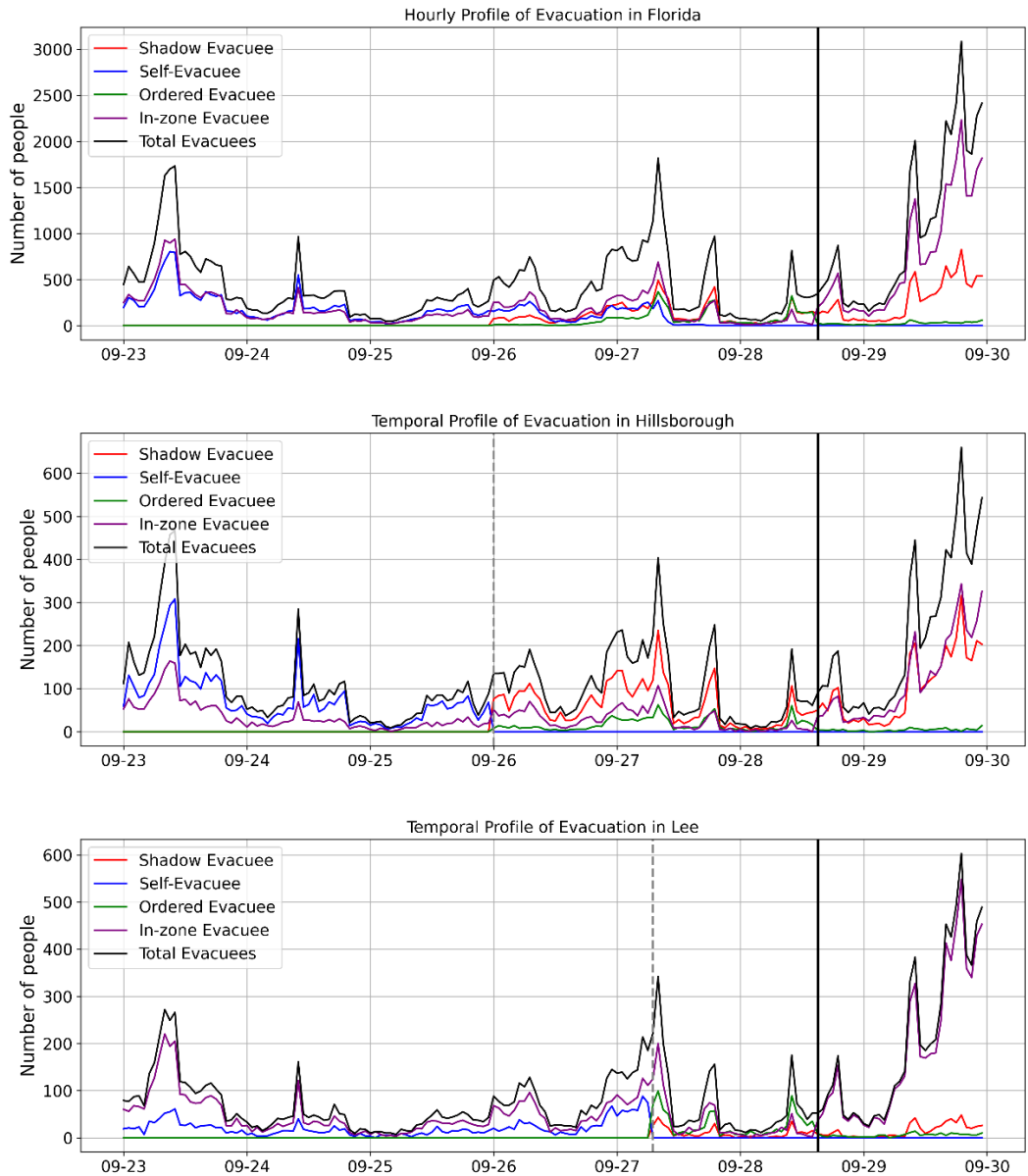

Figure 7: Hourly profile of evacuation rate in all affected areas in Florida, Hillsborough County, and Lee County. Dashed vertical line represents the issue time of the evacuation orders, and black vertical line represents the landfall time.

Figure 8 further offers more nuanced details of the spatial patterns of evacuation rate during the three waves by each users' departure time from home, respectively. Similar to the results shown in Figure 7, the maps show the third wave after the landfall is the largest wave in terms of number of evacuee, and it shows that the hypercautious evacuees who left home way before the hurricane is rather widespread and not limited to a certain area, although they are fewer in number. The maps also reveal an interesting phenomenon that the evacuation rates are becoming more clustered and heterogenous from the first wave to the third wave, and there is a notable increase in evacuation rates near coastal areas during the third wave. This pattern shift suggests an increased urgency, such as infrastructure damage and water or power failure, or a last-minute decision-making process among residents in coastal zones who primarily get affected by the hurricane.

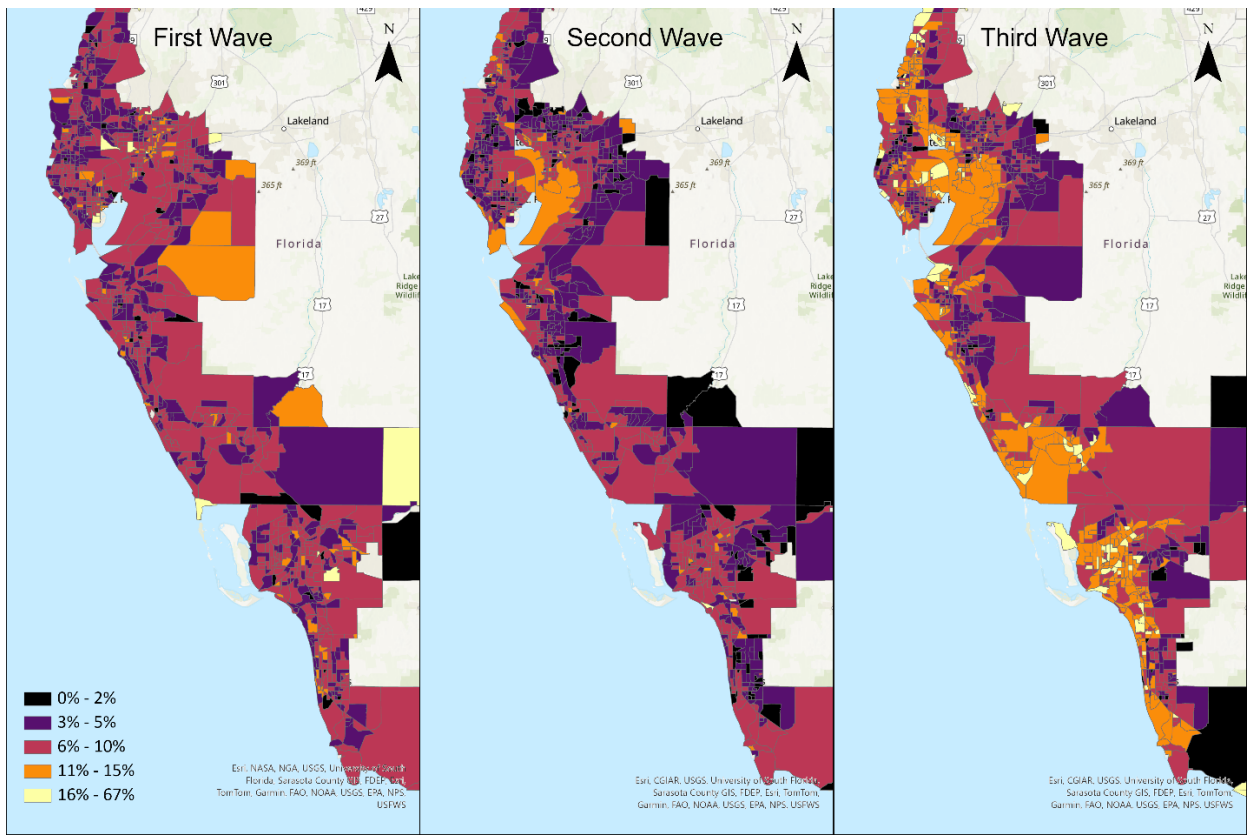

Figure 8: The spatial pattern of evacuation rate during three waves.

## 5. Discussion and Conclusion

With increasing frequency, intensity, and unpredictability of hurricane events under the growing effects of climate change, there is an urgent need to analyze travel behaviors of large-scale evacuations during hurricanes. Hurricane Ian, the deadliest and costliest hurricane in Florida's history and the third costliest in the United State history, further underscores this critical need. The emergence of large-scale mobile device location data offers an unprecedented opportunity to explore these behaviors in detail, providing high-resolution and high-fidelity insights into the spatial and temporal dynamics of evacuee movements. By using a terabyte-level mobile device location dataset, we develop a novel hurricane evacuation behavior algorithm to categorize evacuees into distinct groups based on their mobility patterns and infer their departure times. This algorithm allows for a more nuanced understanding of the decision-making processes during evacuations and can pinpoint critical times and locations where interventions could be most effective.

The spatiotemporal analyses conducted on the GPS data reveals many very useful and practical insights. First, the spatial analysis of evacuation behaviors during Hurricane Ian, as shown in the study, highlights a complex pattern of compliance and mobility across Florida. Despite the critical nature of the situation, evacuation rates were notably lower than in previous hurricanes, reflecting issues with evacuation order timing and public compliance. In Lee County (Cape Coral and Fort Myers), which accounted for most of the casualties, we observe a notably lower out-of-zone evacuation rate, but a higher general evacuation rate, likely due to late evacuation orders and confusion among residents. Conversely, Hillsborough County (Tampa) demonstrated a higher compliance rate, aided by earlier evacuation orders which allowed residents more time to react and prepare. Second, the temporal patterns of evacuation further reveal three primary waves of evacuation, each corresponding to different stages of the hurricane's approach and aftermath. The data reveals that first wave is very early around the formation of the hurricane, followed by a spike just before the hurricane landfall and a significant post-event spike. Finally, although general evacuation rate would decrease from coastal to inland areas, which has extensively reported by prior studies (Mongold et al., 2021; Washington et al., 2024), we find that out-of-zone evacuation rates were actually higher for inland areas, possibility due to geographical proximity and better traffic expectation. These nuanced findings demonstrate the necessity to utilize more high-resolution, high-fidelity, and high-frequency data for conducting in-depth analyses of people's behaviors during crises.

This paper has many practical insights for future disaster planning and policymaking.

- First, many findings in this paper suggest that the timing of evacuation orders made a significant impact on the patterns of evacuation. Our findings show that areas with early evacuation orders have relatively higher full compliance rate. Despite being extensively discussed by prior studies (Han et al., 2007; Wolshon et al., 2005), the case study of Ian further exemplifies the critical importance of timely and clear evacuation order.
- Second, our results show a highly heterogenous patterns in the planning and response of a major impending disaster among different counties and communities, which underscores the need for a coherent tailored communication and planning mechanism at the region or state level. Evacuation zones and evacuation orders are managed independently by each county using different platforms, such as official websites, Facebook, and Twitter (Citrus County, 2022; Highlands County, 2022; Manatee County, 2023). Despite multiple endeavors to organize and communicate these information at state (Florida Division of Emergency Management, n.d.) and federal level (NOAA, 2024), this variability complicates the acquisition of correct and timely evacuation updates, potentially leading to confusion and hesitancy to evacuate. For example, a major portion of the evacuees are shadow evacuees, who lives near but outside the designated evacuation zones.

- Third, the new datasets and methods introduced in this study reveal more nuanced and detailed insights into evacuation behaviors than previously seen in earlier research. With large-scale mobile device location data, our method achieves more fine-grained classification of evacuation users, high-fidelity spatiotemporal pattern, and improved understanding of decision-making during evacuations. These insights not only enhance immediate response efforts but also contribute to long-term policy making and urban planning that prepare communities better for future disasters.

This paper has some limitations and potential directions that future studies can address. First, although we extensively discussed the evacuation behaviors, we do not address the return behaviors. Due to the long duration of a hurricane event, it would be hard to determine a returning trip from the GPS points. Instead, we make sure out-of-zone evacuees have no activities happening inside the evacuation zones, which guarantees the correctness of the inference results. However, it is indeed important to investigate these trips, possibly in tandem with a study of evacuation destination and evacuation routes. Second, we are unable to determine the purpose or motivation behind trips using GPS data alone, as it does not capture the psychological factors influencing people's decisions. Future studies can use survey-based methods to fill in this gap. Finally, despite being a satisfactory proxy in general, mobile device location data can have representativeness issues at smaller spatial scales, leading to overestimation or underestimation of evacuation behaviors for certain neighborhoods (Jardel and Delamater, 2024; Z. Li et al., 2024). Future studies should validate and complement the findings with survey-based results, as GPS data-based inference method provides an incomplete picture of human behavior during disasters.

## Statements

### Declaration of interest

The authors have no financial and personal relationships with other people or organizations that could inappropriately influence the work.

### Author contributions

**Luyu Liu:** Conceptualization; Data curation; Formal analysis; Investigation; Methodology; Project administration; Supervision; Validation; Visualization; Roles/Writing - original draft; and Writing - review & editing.

**Xiaojian Zhang**: Data curation; Investigation; Formal analysis; Methodology; Writing - review & editing.

**Shangkun Jiang:** Formal analysis; Investigation; Methodology.

**Xilei Zhao**: Conceptualization; Data curation; Project administration; Supervision; Validation; Writing - review & editing.

### Funding source

We are grateful for the funding support from the Center for Equitable Transit-Oriented Communities Tier-1 University Transportation Center (Grant No. 69A3552348337).